\documentclass[letterpaper,aps,floatfix,twocolumn]{revtex4}

\usepackage{graphicx}
\usepackage{epsfig}
\usepackage{amsmath,amsthm}
\usepackage{amssymb}
\usepackage{amsfonts}
\usepackage{psfrag,overpic}
\usepackage{mathrsfs}
\usepackage{hyperref}
\usepackage{alltt}
\usepackage{setspace}
\usepackage{color}
\usepackage{dsfont}

\newcommand{\old}[1]{{}}
\def\dd{{\rm d}}

\begin{document}

\title{Maximum Entropy Production as a Necessary Admissibility Condition
for the Fluid Navier-Stokes and Euler Equations}

\author{James Glimm}
\email{glimm@ams.sunysb.edu}
\affiliation{Stony Brook University, Stony Brook, NY 11794, USA, and GlimmAnalytics LLC}
\author{Daniel Lazarev}
\email{daniel.lazarev@stonybrookmedicine.edu}
\affiliation{Stony Brook University, Stony Brook, NY 11794, USA}
\author{Gui-Qiang G. Chen}
\affiliation{University of Oxford, Oxford, OX2 6GG, UK}
\email{chengq@maths.ox.ac.uk}
\date{\today}

\begin{abstract}
In a particle physics dynamics,
we assume a uniform distribution as the physical measure
and a measure-theoretic definition of entropy
on the velocity configuration space.
This distribution is labeled as the physical solution
in the remainder of the article.
The dynamics is governed by an assumption of a Lagrangian formulation,
with the velocity time derivatives as the momenta conjugate to the
velocity configurations.
From these definitions and assumptions, we show
mathematically that a maximum entropy production principle
selects the physical measure
from among alternate solutions of the Navier-Stokes and Euler equations,
but its transformation to an Eulerian frame is not established
here, a topic that will be considered separately.

\end{abstract}
	
\maketitle
\section{Introduction}
\label{sec:intro}

\subsection{Background}
\label{sec:background}
\makeatletter{\renewcommand*{\@makefnmark}{}
\footnotetext{\\ 2020 \textit{Mathematics Subject Classification:} 46A13, 37A50}\makeatother}

Turbulence is perhaps the foremost problem of classical
(nonquantum) physics.
The main result of this paper is a proof of a
maximum rate of entropy production as a necessary admissibility principle
for physical solutions of fluid turbulence.
For a mathematical analysis, as we present here,
we define the physical measure on the fluid velocity
configuration space as proportional to the
uniform distribution.

Our result is a necessary condition for the admissibility of solutions of the
Navier-Stokes and Euler equations of fluid dynamics.
The examples of nonunique solutions of the
Navier-Stokes equations \cite{BucVic19} and the Euler equations
\cite{Sch93a,DeLSze09,BucVic20,GliShaLim15}
show the need for such an admissibility principle.

For fluid mixing, an entropy of mixing, similar in form to
the entropy considered in this paper, but
defined based on concentration gradients,
is subject to the same physical laws of entropy maximization.
The total entropy of a turbulent fluid can be decomposed into a sum
of a configurational or thermal entropy related to fluid density
fluctuations,
an entropy of mixing related to (thermal) concentration gradients, and
a  kinetic energy entropy related to fluid velocity fluctuations. For
a constant density fluid, that is, classical turbulence, as considered here,
the entropy is exclusively kinetic, and related to the fluid velocities.

\begin{list}{}{\leftmargin=\parindent\rightmargin=0pt}
\item
\textbf{Theorem 1.}  The physical measure
maximizes the entropy production rate
in comparison to alternate measures on the
velocity configuration space.
\end{list}

The distinction between Ziegler's principle of maximum entropy production \cite{Zie63,ZieWeh87}
and Prigogine's principle of minimum entropy production \cite{Pri78}
lies in the distinction between open and closed systems.
In an open system, the entropy, which must increase for any irreversible processes,
can escape to the external world and not be present in the model.
For a closed system, there is no such escape.
Thus, we view Ziegler's principle as applicable to closed systems and
Prigogine's principle as applicable
to open ones. To illustrate this distinction, consider the
irreversible experiment of dropping a stone from the tower of Pisa.
The entropy, which must increase due to the irreversibility of the experiment,
is not found in the stone. Considered in isolation, the stone is an open
system and its entropy increase is minimized (zero) according to Prigogine's principle.
The entropy does increase within the air disturbed by the falling stone.
This disturbance accounts for the aerodynamic (turbulent) drag on the falling stone
and, if included in the model, the system
has a maximum rate of increase, according to Ziegler's principle.
Therefore, we see the Prigogine-Ziegler distinction not as a controversy regarding
laws of physics but as alternate modeling strategies in
the construction of a physical model.

All models involve approximations or idealizations.
The selection of a model is a judgment on the part of the
modeler and is not uniquely determined by the problem.
For this reason, the two opposite principles coexist.
In the case of observational or experimental turbulence, the dominant
dissipation and entropy production occur in boundary layers.
Rather than modeling boundary layer entropy production
explicitly, carefully controlled measurements
of turbulence are generally located in regions of space far from such
boundaries.
This convention explains the fact that turbulence measurements generally
support Ziegler's principle and not Prigogine's principle.
Thus, the main result of the program proposed here, in support of Ziegler's principle, is
that classical fluid turbulence is a closed system.

The role of stirring forces brings this distinction into sharper focus.
Stirring forces are governed by ideas of Prigogine, while the non-stirred,
or what we call Never Stirred (NevS) set of initial conditions and of Navier-Stokes solutions 
(which have no memory of past stirring) are governed by Ziegler's rule, 
according to which the physical solution selects the entropy maximizing solution
from among the nonunique solutions of the Navier-Stokes equations. 
This set of initial conditions is time invariant. 
Put differently, the physical solution belongs to a NevS subspace, velocity
configurations stirred neither in the past nor in the present nor in the future.
This is the subspace of Navier-Stokes initial conditions to which
the maximum entropy condition applies. From this subspace, the
general solution, with stirring in the past, present or future is readily
constructed by coordinate transformations, thereby recovering a general solution
from a NevS one.

Maximum entropy production rates and maximum energy dissipation rates are very general
scientific principles, and their importance extends well beyond the single fluid turbulence considered here.
Often interfaces or near interfaces defined by narrow diffusion layers arise in the study of fluid mixtures.
Efforts to improve the modeling of atmospheric turbulent boundary layers can be based on the
maximum dissipation rate ideal emphasized here. Likewise, energy transfer within the Inertial Confinement
Fusion (ICF) capsule is subject to plasma instabilities, likely of a turbulent origin, and the resulting
boundary layer enhanced ablation at the ICF capsule surface that has been observed is likely a turbulent boundary
layer interaction, deleterious to the ICF yield.
Experimental studies of acceleration driven fluid mixing are typically performed either with a sharp
interface (immiscible fluids) or with a nearly immiscible mixture (high Batchelor number).
Thus, we see multiple reasons to extend the present analysis to mixtures and in doing this,
to consider the role interfaces as defining the nature of the mixtures.

\subsection{Prior studies}
\label{sec:prior}

The maximum entropy production principle (MEPP), as a principle of physics,
has a long history, of which we cite \cite{Daf73a,Daf89,MarSel06,MihFarPai17}.
This principle has been used in two key ways:

\noindent
(i) to derive an evolution system of equations;

\noindent
(ii) to select the physically relevant solution of a given evolution system.

\subsubsection{Particle physics}

MEPP has been used to derive laws such as
Kirchhoff's circuit laws \cite{ZupJur04}
and Fourier's law \cite{LucGra15}.
Jordan, Kinderlehrer, and Otto showed that the heat equation
is the gradient flow, or steepest descent, of a functional equal to the negative of the Boltzmann-Gibbs
entropy \cite{JorKinOtt97,JorKinOtt98}.
A proof of entropy maximization for lattice gases is
given by Lanford \cite{Lan73}.

\subsubsection{Navier-Stokes and Euler equations{\rm :} prior results}

As a selection principle, MEPP has been used in a variety of applications, from the study of
enzyme kinetics \cite{DobVitBru17} to the spatial organization of vegetation
in river basins \cite{DelFotRin12}.
In the physical analysis of fluid turbulence, MEPP has been successfully used to select the physical
solution for the evolution of the global climate system \cite{OzaShiSak01,OzaOhmLor03}.
In regard to the Euler equations, successful results have been obtained for the
two-dimensional case by Chavanis et al.  \cite{ChaSomRob96},
who maximized
the mixing entropy to obtain the equilibrium distribution, given standard energy
and momentum constraints, and then used MEPP to derive the equations governing
the evolution of the configuration space density probabilities.
 Boucher et al. \cite{BouEllTur00} derived maximum entropy principles for continuum models
of two-dimensional turbulence by using the theory of large deviations.
Thalabard et al. used entropy maximization to select solutions for the three-dimensional
axi-symmetric Euler equations in a Taylor-Couette geometry \cite{ThaDubBou18}.
Axi-symmetric flow in 3D is also two-dimensional.

\section{\, Properties of the Physical Measure}
\label{Phys-measure-properties}

The physical measure is defined as proportional to the Lebesgue
measure in a framework of particle physics.
We identify two 
basic properties or restrictions on the physical measure. It must

\noindent
(i) be restricted to the entropy of
indistinguishable particles only;

\noindent
(ii) be restricted to the surface of
constant energy.

\smallskip
Both
of these conditions are then imposed on the alternate measures
used for comparison to the physical measure.

\subsection{The Navier-Stokes equations}
\label{sec:NS}

\begin{list}{}{\leftmargin=\parindent\rightmargin=0pt}
\item
{\textbf{Definition 1.}}  Let $V$ be a finite cube in $\mathds{R}^3$,
and $[0,T]$ a finite time interval.
Let ${\cal{H}}$ be the $L_2$ space of divergence free velocity
fields defined over $V$ with periodic boundary condition.
The \textit{distinguishable particle configuration
space} ${\cal{V}}^d(V\times [0,T])$ is the
space ${\cal{M}}([0,T];{\cal{H}})$ of Radon measures of $t \in [0,T]$ with
values in  ${\cal{H}}$.
\end{list}

The physical measure and all comparison measures are realized as
linear functionals defined on the
configuration space ${\cal{V}}^d$.
In view of the self-duality for any Hilbert space ${\cal{H}}$,
we identify the dual
${\cal{M}}([0,T];{\cal{H}}^*)$ as the space ${\cal{C}}([0,T];{\cal{H}})$
of continuous functions on $[0,T]$ with values in ${\cal{H}}$.
We use the weak topology on ${\cal{M}}$ to define the open sets.
The Borel sets in ${\cal{M}}$ are members of the $\sigma$-ring generated by the open sets.

\begin{list}{}{\leftmargin=\parindent\rightmargin=0pt}
\item
{\textbf{Definition 2.}} The physical measure on the configuration
space ${\cal{V}}$ is proportional to the Lebesgue measure.
\end{list}

The physical measure is invariant under translations
by velocities $u \in {\cal{V}}$ and by two independent angular rotations.
The first modification of the physical measure
is the specification of the Navier-Stokes
equations in the Lagrangian frame which it solves.
With a viscosity
$\nu(x,t)$ given, the Lagrangian
Navier-Stokes equations (after projection of the system
onto the divergence free subspace)
at fixed time $t$ is
\begin{equation}
\label{eq:NS}
\frac{D}{Dt} u(x,t) = \nabla\cdot(\nu(x,t)\nabla u(x,t))
\end{equation}
for a divergence free $u(x,t) \in {\cal{H}}$. The nonlinearity in this
evolution system enters through the material derivative $D/Dt$.
As $\nu$ enters in (\ref{eq:NS}) multiplying second velocity derivatives,
it is natural to require $u(x,t) \in {\cal{H}}  _{+2}$,
the Sobolev space of functions with
two $L_2$ spatial derivatives. Since $u$ lies in the dual space
of functions continuous in $t$, we suppose that
$u(x,t) \in C([0,T],{\cal{H}}_{+2})$ of continuous functions of time
with values in ${\cal{H}}_{+2}$. The regularity condition is weakened to
$u \in {\cal{H}}$ through the use of cylinder sets (defined below) with
smooth basis elements, and projective limits.

We impose the
solution restriction (\ref{eq:NS}) of the Navier-Stokes equations
on the comparison
measures to which the physical measure is compared. The validity of
these equations for the physical solution will be addressed separately.

\subsection{Transformation group invariance properties}
\label{sec:Noether}

The subspace of the configuration space ${\cal{V}}$
defined by the Navier-Stokes (or Euler) equations (\ref{eq:NS})
is invariant by elements of
the semidirect product of the group of translations
by a velocity $u(x,t)$ in the subspace defined by the Navier-Stokes (or Euler) equations
(\ref{eq:NS}), and the orthogonal group of transformations
on the vector indices of the velocity.

 The consequence of these symmetries, namely Noether's theorem,
according to which the velocity field $u(x,t)$ in
the Navier-Stokes physical solution subspace of configuration space
${\cal{H}}$ conserves linear momentum and two independent angular momenta as a
thermodynamic identity \cite{Noe1918},
will be established separately. Conservation of linear momentum is equivalent to the Navier-Stokes equation, ie, (\ref{eq:NS}).

\subsection{Indistinguishability}
\label{sec:indist}

We remove the entropy distinction between labels for individual particles.
This goal is achieved through the definitions of
distinguishable and indistinguishable configuration spaces and the
linear functionals on them. To identify individual particles
(distinguished or not) within the particle physics formalism, we introduce
cylinder sets defined by finite dimensional subspaces ${\cal{H}}_n$ of
${\cal{H}}$.
To illustrate the ideas, we work out the details explicitly for $n = 2$
particles
at a fixed time $t \in [0,T]$. The two particle configuration space
${\cal{H}}_2$ is the set of
2-tuples, $(u_1(x_1),u_2(x_2)) \in {\cal{H}} \times {\cal{H}}$.
If the particles are distinguishable,
then states $(u_1(x_1),u_2(x_2))$ and $(u_2(x_2),u_1(x_1))$ are distinct.
If they are
indistinguishable, we consider the symmetry group ${\cal{S}}_2$ on two
objects, and regard the orbits under ${\cal{S}}_2$ acting on ${\cal{H}}_2$
as the elementary points in the indistinguishable 2 particle configuration space.
Given a measure $\mu_2^d$ on the distinguishable 2 particle configuration space,
we construct an indistinguishable measure $\mu_2$ on the indistinguishable
configuration space by the formula
\begin{equation}
\label{eq:2-part}
\mu_2 = \frac{1}{2!} \mu_2^d
= \frac{1}{|{\cal{S}}_2|} \sum_{s_2 \in {\cal{S}}_2} \mu_2^ds_2(u_1(x_1), u_2(x_2))
\end{equation}
where $|{\cal{S}}_2| = 2!$ is the number of elements in ${\cal{S}}_2$.
Thus, the division by $2!$ is identical to taking the average of the
distinguishable measure of the members of the orbit, averaged
over the orbit. These definitions and calculations
generalize to ${\cal{H}}_n$ and ${\cal{S}}_n$.

We assume that a distinguished basis has been chosen for
${\cal{H}}$, and let ${\cal{H}}_n$ be the span of the first $n$ basis elements.
Using this (self-dual) space, we define the cylinder sets which allow us to
define distinguishability and indistinguishability.

We consider a fixed time $t \in [0,T]$ and define the fixed time cylinder set
$(\delta_t, {\cal{H}}_n) \subset {\cal{M}}([0,T];{\cal{H}})$
of configuration space and its dual
$(t, {\cal{H}}_n) \subset (t,{\cal{H}})$ of functionals on
the configuration space.
Any Borel set $A_n \subset {\cal{H}}_n$ similarly defines
a Borel cylinder set at fixed time.

With this preparation, we define the subspaces of
totally distinguishable and totally indistinguishable configuration spaces
and the linear functionals on these configuration space.
Relative to the basis for ${\cal{H}}$, consider the symmetry group
${\cal{S}}_n$ on $n$ objects, acting on the first $n$
coefficients of the velocity when expanded in this
chosen basis. An element of
${\cal{H}}_n$ is totally distinguishable, while its orbit under
${\cal{S}}_n$ has $n!$ elements and is totally indistinguishable.
These definitions also apply to functionals on the configuration space.

Given a totally distinguishable linear functional $\mu^d$ on the
$n$ particle fixed time configuration space, we associate to it a new
indistinguishable functional $\mu$ on the indistinguishable particle configuration space,
as the average of $\mu^d$ over the symmetry group orbit.
This average is the original measure of the orbit (the sum over the $n!$
orbit elements) divided by $n!$.
In the case of subspace inclusion ${\cal{H}}_m \subset {\cal{H}}_n$,
consistency requirements involving the $n!$ and $m!$ are readily verified.
These same consistency relations are the formal definition of projective
limits for groups, so that an infinite dimensional symmetry
group is defined on the configuration space and linear functionals
on the configuration space. The consistent linear functionals $\mu$ satisfy
the criteria of a projective limit and define a finitely additive fixed time
linear functional on the configuration space.

We require the basis elements in ${\cal{H}}_n$ to be smooth
functions of $x \in V$, allowing spatial derivatives to be evaluated
within cylinder sets.

We use the notion of projective limit to pass from cylinder sets to an infinite space limit \cite{Rao71,Bou17}.

\begin{list}{}{\leftmargin=\parindent\rightmargin=0pt}
\item
{\textbf{Definition 3.}}
Consider a projective system $(X_n, g_{mn})$ defined by an object $X_n$
and maps $g_{mn}: X_n \rightarrow X_m$
satisfying the consistency condition:
\begin{equation*}
\label{eq:consis-proj}
g_{lm} \circ g_{mn} = g_{ln} \qquad \mbox{for $l \leq m \leq n$}.
\end{equation*}
Then the \textit{projective limit} is $(X, g_n)$,
where $X = \displaystyle \lim_{\longleftarrow}X_n$ is an object and $g_n$ is a map $g_n: X \rightarrow X_n$ satisfying:
\begin{equation*}
\label{eq:consis-proj2}
g_{m} = g_{mn} \circ g_{n}.
\end{equation*}
$(X,g_n)$ is universal if, for any other pair $(Y,f_n)$ with $f_n: Y \rightarrow X_n$ satisfying [\ref{eq:consis-proj2}],
there is a unique morphism $k: Y \rightarrow X$ such that $f_n=g_n \circ k$.
\end{list}

\begin{list}{}{\leftmargin=\parindent\rightmargin=0pt}
\item
{\textbf{Proposition 1.}} The system $X$ consists of the following structures: 
\begin{enumerate}
\item[\rm (i)]  $(t,{\cal{H}}_n), t \in [0,T]$,

\item[\rm (ii)] $A_{\rm Borel} \subset {\cal{H}}_n$, 

\item[\rm (iii)] ${\cal{S}}_n$, 

\item[\rm (iv)] indistinguishable particle configuration space and linear functionals $\mu_n$.
\end{enumerate}

The consistency conditions of Definition 3 are satisfied, as are the
the universality conditions of the projective limit thus defined.
\end{list}

\textit{Proof.} We consider the morphisms for each of the cases individually: 

\smallskip
(i) $g^{-1}_{mn}:{\cal{H}}_m \rightarrow {\cal{H}}_n$ is the inclusion map of a smaller Hilbert space to a larger one.  

\smallskip
(ii) $g^{-1}_{mn}$ is an inclusion map.

\smallskip
(iii) $g^{-1}_{mn}:{\cal{S}}_m \rightarrow {\cal{S}}_n$ is the inclusion map from a smaller symmetry group to a larger one. 

\smallskip
It is readily verified that (i)--(iii) satisfy the necessary consistency conditions.

\smallskip
(iv) Let ${\cal{V}}_l$, ${\cal{V}}_m$, and ${\cal{V}}_n$ be the indistinguishable cylinder subsets of ${\cal{V}}$
such that ${\cal{V}}_l \subset {\cal{V}}_m \subset {\cal{V}}_n$,
and let $\mu_l$, $\mu_m$, and $\mu_n$ be the measures defined on them, respectively.

Without loss of generality, we let their dimensions (denoted by their respective subscripts) satisfy $l<m<n$.
Define a restriction map $g_{mn}:  {\cal{V}}_n \rightarrow  {\cal{V}}_m$  given by $g_{mn}(A_n)=\frac{m!}{n!} A_m$,
where $A_l \subseteq  {\cal{V}}_l$, $A_m \subseteq  {\cal{V}}_m$, $A_n \subseteq  {\cal{V}}_n$,
and $\mu_l^d(A_l)=\mu_m^d(A_m) = \mu_n^d(A_n)$.
Then
\begin{align*}
\nonumber
 g_{lm} \circ g_{mn}(A_n) &= g_{lm}\left(\frac{m!}{n!} A_m \right) \\
 &=\frac{l!}{m!}\frac{m!}{n!} A_l \\
 &= \frac{l!}{n!} A_l \\
 &= g_{ln}(A_n).
\end{align*}

The linear functionals on the indistinguishable particle configuration space equal
those on the distinguishable particle configuration space, after division by a factor of $n!$.
Given the restriction map $g_{mn}$, we verify that these indistinguishable particle linear functionals
are consistent:
\begin{align*}
\mu_m \circ g_{mn} (A_n) &= \mu_m \left(\frac{m!}{n!} A_m \right)\\
   &= \frac{1}{m!} \frac{m!}{n!} \mu_m^d(A_m) \\
&= \frac{1}{n!} \mu_n^d(A_n)\\
&= \mu_n (A_n).
\end{align*}

We next prove the universality of the projective limit.
Consider the projective limits $P_1$ and $P_2$,
each defined by an ordered basis set
such that $P_1 \subset P_2$ if every basis element of $P_1$ occurs
within the $P_2$ basis list.

We verify the conditions for $g_{nm}^{-1}$ rather than
those for $g_{mn}$.
These are simply subset relations and are trivially
consistent, other than the $n!$ analysis for indistinguishable objects,
which are verified as before.

If $P_1 \subset P_2$ and $P_2 \subset P_1$,
then the two basis sets are identical up to reordering.
In this case, we have the identity of the projective limits, so that we write $P_1 = P_2$.

Given a general $P_1$ and $P_2$, neither subsets of the other, we consider the projective limit ${\cal{P}}$
which chooses basis elements alternately from the $P_1$ and $P_2$ basis sets.
Then $P_1 \subset {\cal{P}}$ and $P_2 \subset {\cal{P}}$, and each is a subset of the projective limit ${\cal{P}}$.

The universal limit ${\cal{P}}$ is the union over all possible projective limits $P_i$,
and it is universal in that any projective limit is a subset of ${\cal{P}}$.
\hspace*{\fill} $\square$ \\

Projective limits are used to define the physical (Lebesgue) measure on the configuration space of distinguishable velocity values,
the symmetry group of velocity interchanges, the configuration space of indistinguishable velocity values
(which is the quotient space of equivalence classes under this symmetry group),
and the physical (Lebesgue) measure on the indistinguishable velocity configuration space.

Limits of consistent measures to a finitely additive limit are known as a special case of this formalism.
The formalism serves mainly to clarify the conceptual issues relating to the limit over cylinder sets.
We proceed at the level of cylinder sets, where most of the analysis of the paper occurs.

\begin{list}{}{\leftmargin=\parindent\rightmargin=0pt}
\item
{\textbf{Definition 4.}} The \textit{physical measure $\mu_n^{d,*}$ on an $n$-dimensional cylinder subset
of the distinguishable configuration space} is defined by the Lebesgue measure on that cylinder subset.

The \textit{physical measure $\mu^{d,*}$ on the distinguishable particle configuration space} is
the finitely additive projective limit of the projective system defined by the cylinder subsets
and the physical cylinder set measures on them.
\end{list}

The factor $n!$ accounts for the fact that the interchange of velocities $u_i \leftrightarrow u_j$ defines
an identical point in the configuration space of indistinguishable particles.

\begin{list}{}{\leftmargin=\parindent\rightmargin=0pt}
\item
{\textbf{Definition 5.}} The \textit{indistinguishable particle configuration space of order $n$},
${\cal{V}}_n$, is the set of equivalence classes of the cylinder set configuration space of dimension $n$
under the particle interchange symmetry.
The indistinguishable physical measure $\mu_n^* = \mu_n^{d,*}/n!$ is the Lebesgue measure divided by $n!$.

The \textit{physical measure $\mu^*$  on the indistinguishable particle configuration space} is
the finitely additive projective limit of the projective system defined by the cylinder subsets
and the physical cylinder set measures on them.
\end{list}

\subsection{The energy surface}

Thermodynamically, entropy is a function of two thermodynamic
variables, density $\rho$ and energy $e$. These can be specified in
a variety of ways, depending on the thermodynamic ensemble assumed.
In the microcanonical ensemble, both $\rho$ and $e$ have prescribed values.
\label{sec:eng-surf}
We set the
density $\rho = 1$, so that the energy
$e$ is given by $e(u)(x,t) = (u^2/2)(x,t)=u(x,t)^2/2$. We specify
$e(u)(t) = \|e(u)(\cdot,t)\|_{L_1(V)}$
as defining the fixed time energy
surface to be imposed as a constraint on the configuration space.

\begin{list}{}{\leftmargin=\parindent\rightmargin=0pt}
\item
{\textbf{Definition 6.}}  A \textit{$($fixed time$)$ candidate measure $\mu_n$ on the indistinguishable particle configuration space}
is a consistent family of  indistinguishable particle
Radon measures defined on the cylinder set configuration space  ${\cal{V}}_n$  such that each cylinder set measure is
absolutely continuous with respect to the (fixed time) indistinguishable
particle Lebesgue measure. The measure is assumed to be restricted to the
energy surface and to be invariant under the orthogonal group and translations, both
acting on the velocities.

\end{list}
The physical measure at fixed time is  a candidate measure.

\begin{list}{}{\leftmargin=\parindent\rightmargin=0pt}
\item
\textbf{Proposition 2.}
Fix a value of $t \in [0,T]$, and let $\mu_n$ be a candidate measure on the
distinguishable particle cylinder set configuration space.
Let ${\cal{V}}_n(e,V)$ denote the energy surface of the cylinder subset of the configuration space
with  $r=r(e)$ its radius, and let
$f=f(u)$ be a continuous functional of the velocities $u$ in ${\cal{V}}_n(e,V)$.
The restriction of the Radon measure $\mu$ to
$e(u)(t)$ is defined as the measure $\mu_e$ on ${\cal {V}}_n(e,V)$ given by
 \begin{equation*}
 \label{eq:eng-surface}
 \int f \left |_{_{e(u) = e}}  \dd \mu_{n,e}
 = \left (\dfrac{\dd}{\dd r}\int f\,\dd\mu_n \right )\right |_{r = r(e)}.
 \end{equation*}
\end{list}
In other words, on the left-hand side,
$f$ is restricted to the energy surface and, on the right-hand side,
the integral of $f$ is evaluated at the same value of $e$.

\textit{Proof.} We change variables to a radial variable and $n-1$
angular variables. The Jacobian for this transformation is smooth away
for $r = 0$. In these coordinates, the evaluation of the integral is elementary.
\hspace*{\fill} $\square$

\section{\, Measure Theoretic Entropy}
\label{sec:info-ent}

The thermodynamic entropy is defined by Boltzmann and Gibbs as
\begin{equation}
\label{eq:bgs-entropy}
S_{BGS} (\sigma,{\cal{V}}_n(e,V))
= - \int_{{\cal{V}}_n(e,V)} \sigma  \log (\sigma)\;\dd^3v,
\end{equation}
where $\sigma$ is a probability density function  \cite{Whe91,DunTalHan07,GavCheBec17}.
The information-theoretic entropy defined by Shannon for discrete random variables
is extended to continuous random variables by a similar formula \cite{CovTho91}.
A generalization of these definitions is the Baron-Jauch entropy \cite{Whe91,BarJau72,Ska75}.
\begin{list}{}{\leftmargin=\parindent\rightmargin=0pt}
\item
\textbf{Definition 7.} The (Baron-Jauch) \textit{entropy}
defined by a  probability  measure $\mu_e$ on a measure space
$\mathfrak{V}$ relative to a  measure $v_e$
on $\mathfrak{V}$, (where $\mu_e$ is absolutely continuous with respect to $v_e$)
is
\begin{equation}
\label{eq:e-working-def}
S(\mu_e,\mathfrak{V})
= -\int_{\mathfrak{V}}\left(\dfrac{\dd\mu_e}{\dd v_e} \right)  \log \left(\dfrac{\dd\mu_e}{\dd v_e} \right) \; \dd v_e.
\end{equation}
\end{list}

In what follows, we take $\mathfrak{V}={\cal{V}}_n(e,V)$,
the cylinder subset of the configuration space ${\cal{V}}$ restricted to the energy surface, and use the corresponding cylinder set measure.
This entropy depends on the reference measure $v_e$.
It was shown \cite{DunTalHan07}
that the Baron-Jauch entropy (\ref{eq:e-working-def}) agrees with
the thermodynamic entropy (\ref{eq:bgs-entropy})
when the reference measure is the Lebesgue measure.
Thus, we let $v_e$ be the indistinguishable particle Lebesgue measure restricted to the energy surface.

\begin{list}{}{\leftmargin=\parindent\rightmargin=0pt}
\item
\textbf{Remark 1.} The formula (\ref{eq:e-working-def})
can be extended to define the entropy for a finite but not necessarily unitary measure restricted to the energy surface.
By defining the probability measure in (\ref{eq:e-working-def})
as
$$
\mu_e(X) = \frac{\eta_e(X)}{\eta_e({\cal{V}}_n(e,V))},
$$
with $\eta_e(X)$ being a finite measure, we obtain
\begin{align}
&S(\eta_e,{\cal{V}}_n(e,V))\nonumber\\[1mm]
&= \log \eta_e({\cal{V}}_n(e,V)) \nonumber\\
&\quad - \frac{1}{\eta_e({\cal{V}}_n(e,V))} \int_{{\cal{V}}_n(e,V)}\left(\dfrac{\dd\eta_e}{\dd v_e} \right)
 \log \left(\dfrac{\dd\eta_e}{\dd v_e} \right)  \dd v_e. \label{eq:e-true-def}
\end{align}
\end{list}
All our results below hold for finite measures by using (\ref{eq:e-true-def}) instead of (\ref{eq:e-working-def}) in the proofs.
However, proceeding with (\ref{eq:e-working-def}) and probability measures allows for cleaner, less congested proofs.
Using (\ref{eq:e-true-def}) in the proofs,
one can readily verify that all our results for the physical measure are obtained for any measure proportional
to the Lebesgue measure ({\it i.e.}, for the Lebesgue measure with a constant prefactor).

\section{\, Maximum Entropy and Admissibility}
\label{sec:entropy1}

A key step in the proof of our main theorem, Theorem 1, is a reformulation of
the entropy defined by the physical measure.

 \begin{list}{}{\leftmargin=\parindent\rightmargin=0pt}
\item
 \textbf{Proposition 3.}
\begin{equation}
\label{eq:equal-defs*}
 S(\mu_{n,e}^*,{\cal{V}}_n(e,V)) = \log|{\cal{V}}_n(e,V)|,
\end{equation}
where $| \cdot | = v_e(\cdot)$ denotes the indistinguishable particle Lebesgue measure
restricted to the energy surface.
\end{list}

\textit{Proof.} We write $\mu_{n,e}^*(X) = |X| / |{\cal{V}}_n(e,V)|$.
Then its Radon-Nikodym derivative is
$$
\frac{\dd\mu_{n,e}^*}{\dd v_e}=\frac{1}{|{\cal{V}}_n(e,V)|} \ .
$$
Starting with (\ref{eq:e-working-def}) with $\mu_{n,e} = \mu_{n,e}^*$, we have
\begin{align*}
\nonumber
&S(\mu_{n,e}^*,{\cal{V}}_n(e,V)) \\
&= -\int_{{\cal{V}}_n(e,V)}\left(\dfrac{\dd\mu^*_{n,e}}{\dd v_e} \right)  \log \left(\dfrac{\dd\mu^*_{n,e}}{\dd v_e} \right) \; \dd v_e \\
\nonumber
&=-\int_{{\cal{V}}_n(e,V)}\left(\dfrac{1}{|{\cal{V}}_n(e,V)|} \right)  \log \left(\dfrac{1}{|{\cal{V}}_n(e,V)|} \right) \; \dd v_e\\
\nonumber
&=\dfrac{1 }{|{\cal{V}}_n(e,V)|} \log |{\cal{V}}_n(e,V)| \int_{{\cal{V}}_n(e,V)}  \dd v_e\\
&= \log |{\cal{V}}_n(e,V)|,
\end{align*}
which completes the proof. \hspace*{\fill} $\square$ \\

We show that entropy production is maximized by the physical measure defined on a cylinder subset of
the configuration space restricted to the energy surface.
\begin{list}{}{\leftmargin=\parindent\rightmargin=0pt}
\item
\textbf{Proposition 4.}
The entropy production on the cylinder sets of the configuration space  restricted to the energy surface is maximized by the physical measure:
\begin{equation}
\label{eq:thm-1}
 S(\mu_{n,e},{\cal{V}}_n(e,V)) \leq  S(\mu_{n,e}^*,{\cal{V}}_n(e,V)).
\end{equation}
Maximization for each value of $t$ maximizes, in turn, the entropy production rate.
\end{list}

\textit{Proof.} We first maximize the entropy production.
Starting from Eq. (\ref{eq:e-working-def}), we denote $\mu'_e=\dd\mu_e/\dd v_e$
and use the concavity of the logarithm and Jensen's inequality:
\begin{align*}
\nonumber
S(\mu_{n,e},{\cal{V}}_n(e,V)) &=\int_{{\cal{V}}_n(e,V)}\mu'_{n,e} \log \left(\dfrac{1}{\mu'_{n,e}} \right) \; \dd v_e \\
\nonumber
&\leq  \log \int_{{\cal{V}}_n(e,V)} \mu'_{n,e} \left(\dfrac{1}{ \mu'_{n,e}} \right) \dd v_e\\
\nonumber
&= \log |{\cal{V}}_n(e,V)| \\
&= S(\mu_{n,e}^*,{\cal{V}}_n(e,V)),
\end{align*}
where the last equality is due to Proposition 3.
As our analysis proceeds through fixed time cylinder sets,
the maximization holds for each fixed value of $t$ and thus applies to the entropy production rate.
\hfill $\square$ \\

We generalize and strengthen the result of Proposition 4 to the full configuration space restricted to
the energy surface, ${\cal{V}}(e,V)$.

\medskip
\begin{list}{}{\leftmargin=\parindent\rightmargin=0pt}
\item
\textbf{Proposition 5.}
The entropy production on the configuration space restricted to the energy surface is maximized by the physical measure:
\begin{equation}
\label{eq:thm-1a}
 S(\mu_{e},{\cal{V}}(e,V)) \leq  S(\mu_{e}^*,{\cal{V}}(e,V)).
\end{equation}
Maximization for each value of $t$ maximizes, in turn, the entropy production rate.
\end{list}

\textit{Proof.} We analyze at each fixed value of $t$.
Replacing ${\cal{V}}_2$ with ${\cal{V}}_1 \cup {\cal{V}}_2$, etc., we can assume that the ${\cal{V}}_i$'s are a nested sequence of cylinder sets.

We enclose the finite-energy velocity space in a box $V_v$.
The cylinder set configuration space is an $n$-dimensional space,
whose Lebesgue measure is given by
\begin{equation*}
\label{eq:2N-space}
|{\cal{V}}_{n}| \leq \frac{ |V_v|^n}{n!},
\end{equation*}
the inequality reflecting the fact that the box $V_v$ may  be larger
than the velocity space it encloses.

The full configuration space ${\cal{V}}$ is given by the union of the cylinder sets,
which we show to have an upper bound:
\begin{align}
\label{eq:upper-bound}
|{\cal{V}}| =  \bigg\lvert \bigcup_{n=0} ^\infty {\cal{V}}_{n} \bigg\rvert
\leq  \sum_n |{\cal{V}}_{n}|  \leq \sum_n \frac{|V_v|^n}{n!}=e^{ |V_v|}.
\end{align}
We take the logarithm of both sides of (\ref{eq:upper-bound})
(recognizing that, by Proposition 3, the left-hand side is
the physical measure entropy of the full configuration space)
and obtain the  upper bound for $S^{*, \infty}=S(\mu_e^*, {\cal{V}})$:
\begin{equation*}
\label{eq:S-bound}
S(\mu_e^*, {\cal{V}})= \log |{\cal{V}}|  \leq |V_v|.
\end{equation*}
The entropy $S({\cal{V}}_i)=S(\mu_e, {\cal{V}}_i)$ is a monotone increasing
functional by formula [\ref{eq:e-true-def}].
With $S({\cal{V}}_i) \nearrow S^\infty$,
$S^*({\cal{V}}_i) \nearrow S^{*, \infty}$,
and by Proposition 4,  $S({\cal{V}}_i) \leq S^*({\cal{V}}_i)$,
we conclude that  $S^\infty \leq S^{*, \infty}$.
\hspace*{\fill} $\square$ \\

\section{\, Conclusions}
\label{sec:conclusions}

We have proved mathematically that maximum entropy production
is a necessary condition to select the physical solution of the Navier-Stokes and Euler equations for incompressible fluids.
Experimental validation of the laws of turbulence has focused on other issues,
such as the detailed description of multifractal theories.
A refining of this body of work might suffice to confirm this dynamic extension of the second law.
Future mathematical analysis is also required to extend the result, {\it e.g.}, to fluid mixing through
the transport equation or more general physics.
\bibliography{refs}

\end{document}